\documentclass[sigplan,screen,10pt]{acmart}

\usepackage[english]{babel}
\usepackage{makecell}
\usepackage{graphicx}
\usepackage{microtype}

\usepackage{fontawesome5}

\usepackage{todonotes}

\setcopyright{cc}
\setcctype{by}
\acmDOI{10.1145/3759429.3762620}
\acmYear{2025}
\copyrightyear{2025}
\acmISBN{979-8-4007-2151-9/25/10}
\acmConference[Onward! '25]{Proceedings of the 2025 ACM SIGPLAN International Symposium on New Ideas, New Paradigms, and Reflections on Programming and Software}{October 12--18, 2025}{Singapore, Singapore}
\acmBooktitle{Proceedings of the 2025 ACM SIGPLAN International Symposium on New Ideas, New Paradigms, and Reflections on Programming and Software (Onward! '25), October 12--18, 2025, Singapore, Singapore}
\received{2025-04-23}
\received[accepted]{2025-08-11}

\ccsdesc[500]{Software and its engineering~Software architectures}
\ccsdesc[500]{Computing methodologies~Natural language generation}

\keywords{GenAI, software systems, reliability, excellence, design principles, architectural patterns, best practices}

\begin{document}
\title[Foundational Design Principles and Patterns for Building Robust and ...]{Foundational Design Principles and Patterns for Building Robust and Adaptive GenAI-Native Systems}

\author{Frederik Vandeputte}

\email{Frederik.Vandeputte@nokia-bell-labs.com}
\orcid{0009-0001-7471-1345}
\affiliation{
  \institution{Nokia Bell Labs}
  \country{Belgium}
}

\begin{abstract} 
Generative AI (GenAI) has emerged as a transformative technology, demonstrating remarkable capabilities across diverse application domains. However, GenAI faces several major challenges in developing reliable and efficient GenAI-empowered systems due to its unpredictability and inefficiency. This paper advocates for a paradigm shift: future GenAI-native systems should integrate GenAI's cognitive capabilities with traditional software engineering principles to create robust, adaptive, and efficient systems.

We introduce foundational GenAI-native design principles centered around five key pillars---reliability, excellence, evolvability, self-reliance, and assurance---and propose architectural patterns such as GenAI-native cells, organic substrates, and programmable routers to guide the creation of resilient and self-evolving systems. Additionally, we outline the key ingredients of a GenAI-native software stack and discuss the impact of these systems from technical, user adoption, economic, and legal perspectives, underscoring the need for further validation and experimentation. Our work aims to inspire future research and encourage relevant communities to implement and refine this conceptual framework.
 
\end{abstract}

\maketitle

\section{Introduction}

In recent years, Generative AI (GenAI) has demonstrated remarkable emergent capabilities across a diverse array of application domains and use cases. 
GenAI distinguishes itself from traditional algorithms and AI/ML models through its ability to flexibly adopt and apply custom domain knowledge, skills, and reasoning strategies to solve new tasks with minimal additional effort from developers or end users. 
As a result, GenAI is emerging as a versatile and adaptive technology that will profoundly impact future technology stacks and revolutionize our way of working and creativity.

However, GenAI has several key drawbacks compared to traditional algorithmic processing, most notably its unreliability: it is prone to hallucinate despite countermeasures~\cite{huang2025survey, zhang2023siren}, it can be quite unpredictable~\cite{xu2025large, arvan2024interpretability, cao2024worst}, and it has limitations in acquiring new knowledge and skills through prompt engineering~\cite{huckle2025easy, sahoo2024systematic, zhang2023large}.  Other key drawbacks of GenAI, compared to traditional processing, are its large runtime overhead, footprint, and limited debuggability. 

In recent years, several techniques have been developed to help overcome and mitigate the impact of these limitations. These include advanced retrieval augmented generation (RAG) based techniques~\cite{lewis2020rag, gao2023ragsurvey} and collaborative multi-agent systems (MAS)~\cite{liang2023multiagent, hong2023metagpt, guo2024mas, han2024mas} to enhance the knowledge and capabilities of AI agents~\cite{schick2023toolformer}, reasoning and self-reflection~\cite{shinn2023reflexion} capabilities to verify and improve their responses generated through chain-of-thought (CoT)~\cite{wei2022cot} and reasoning~\cite{openai2024learning, guo2025deepseek}, interpretability and explainability techniques~\cite{zhao2024explainability, luo2024understanding} to measure overall quality and uncertainty, reinforcement learning techniques to systematically improve overall end-to-end effectiveness~\cite{chen2024learning, hsu2024grounding, chen2025improving}, etc.

Despite these innovations, GenAI, as cognitive processing technology, will always exhibit some degree of unpredictability, not only due to its architecture---autoregressive, diffusion or neuro-symbolic---but also the inherently dynamic and often under-specified nature of inputs and tasks. 
Moreover, GenAI can be very inefficient compared to traditional algorithms (e.g., calculating 1+1), particularly when utilizing advanced prompting or reasoning techniques.

So instead of concentrating solely on enhancing GenAI technologies, we propose embracing their unpredictability. This unpredictability is a core characteristic of GenAI, allowing it to adaptively generate new, albeit a untested, solutions to both existing and emerging challenges.  Furthermore, we advocate for integrating GenAI approaches with established traditional Software Engineering (SE) methods to develop a balanced, synergetic solution. This may be achieved by designing a self-improving system where GenAI agents systematically automate themselves out of common critical paths.

This requires rethinking and expanding upon the existing software design and development paradigms. By embracing and leveraging their inherent unpredictability and adaptability in addressing new problems, while simultaneously striving for operational efficiency, we can achieve a dual focus on managing uncertainty and enhancing performance. This approach will facilitate the creation of more resilient, adaptive and efficient \textit{GenAI-native systems}.

This paper advocates for a paradigm shift in the development of GenAI-based systems, embracing their limitations while enhancing them with traditional paradigms. 
Through selected use cases, and by drawing insights from historical insights and human methodologies, we introduce a comprehensive conceptual framework of GenAI-native design principles, best practices, and architectural patterns. These include the GenAI-native cell, programmable router, unified conversational interface, and organic substrate. We explore the impact of GenAI-native systems on the future software stacks, and highlight their implications across technological, user, economic, and legal dimensions. Our work aims to redefine future software systems, offering a blueprint for creating resilient, adaptive, and efficient GenAI-native systems, and inspiring future research and innovations. 

\vspace{-4mm}
\section{Motivation}

In this paper, we present a vision for a GenAI-native system as a paradigm that leverages the cognitive capabilities and acknowledges the limitations of GenAI, while seamlessly integrating them with the operational excellence, reliability, and dependability of traditional SE paradigms. Although existing software engineering paradigms provide a solid foundation, we argue that they are inadequate for developing robust and adaptive GenAI-native software systems. On the other hand, \textit{GenAI-first (agentic) systems}, where AI agents manage most critical actions, often lack the robustness and operational efficiency of traditional systems. With \textit{GenAI-native systems}, we aim to combine the strengths of both approaches.

Existing software engineering paradigms focus on designing robust applications with rigid data structures, APIs, and user interfaces. This often results in tightly integrated micro-service architectures. Moreover, in traditional approaches, including low-code and no-code programming, flexibility within functional components is typically preconfigured. In contrast, GenAI offers a more flexible approach, driven by data and reasoning-based methods, potentially using natural language or pseudocode instructions.

As we will discuss, cloud-native principles like immutable infrastructures, CI/CD, and version control need to be  extended and expanded to accommodate to the organic and self-improving nature of GenAI. GenAI allows applications to change its functionality on the fly by integrating custom-generated code, blurring the lines between development and deployment. However, it is crucial to preserve the reproducibility of such assets, and implement restrictions to maintain their original intent, preventing unintended evolution.

The importance of core software engineering principles such as automated testing, monitoring, security, and operational efficiency will only increase and must be further enhanced. 
GenAI-first agentic approaches often lack robustness and operational efficiency, facing challenges like erroneous outputs, unpredictable inconsistencies, and variability in generated responses. Despite efforts to mitigate these issues, the creative nature of GenAI suggests these challenges will persist, especially beyond controlled tests. In addition, operational efficiency will be a key concern, as GenAI-first approaches typically incur higher operational costs with higher processing latency compared to traditional software solutions, mostly due to the complexity of using large models, retrieval augmentation methods, or extended reasoning.

Finally, when developing GenAI-native systems, it is crucial to avoid anthropomorphic pitfalls. While evaluating agents' effectiveness in mimicking human behavior is useful~\cite{openai2025operator, nakano2021webgpt, mialon2023gaia}, these native digital cognitive entities do not require human-oriented interfaces for interacting with web services. GenAI-native systems should be reimagined to allow agents or other GenAI systems to directly communicate with them in a flexible, efficient, and reliable manner.

\vspace{-3mm}
\section{Related Work}
\label{sec:related_work}

While current GenAI-first agentic solutions somewhat reflect the old artisanal methods reminiscent of the early pre-indus\-trial era, we envision a GenAI-native industrial future, where self-reliant multi-agent systems collectively and continuously strive to automate themselves out of the critical path of any solution, yet remain omnipresent to monitor, optimize and help create bespoke solutions.

There is recent work to enhance the robustness and operational efficiency of GenAI-based solutions, which underscore some of the key messages presented in this paper. Code-ge\-ne\-ra\-ting agents generate and execute code snippets instead of using lengthy verbal chain-of-thoughts~\cite{wei2022cot} (e.g., PAL~\cite{gao2023pal}, PAR~\cite{kabra2023par}) or triggering and handling many individual tool calls (e.g. CodeAct~\cite{wang2024codeact}).  Ideally, these generated code snippets should be curated, battle tested, and stored as robust, reusable and efficient solutions for specific inputs (cfr. Dynasaur~\cite{nguyen2024dynasaur}), instead of always requiring agents to reinvent the wheel and regenerate very similar untested solutions. 

The Agora protocol~\cite{marro2024scalablecommunicationprotocolnetworks} is designed to optimize agent-to-agent (A2A) communication~\cite{a2aprotocol2025} by minimizing unstructured, ambiguous natural language exchanges between frequently interacting agents. It facilitates the implementation of a custom (REST) protocol, allowing agents to communicate effectively via a generated functional code implementation. This significantly reduces token usage and processing latency, and results in robust, repeatable, and efficient interactions. Moreover, the protocol retains flexibility, allowing agents to renegotiate or develop additional protocols for diverse scenarios. In addition, the model context protocol (MCP)~\cite{anthropic2024mcp} facilitates seamless integration between large language model (LLM) applications and external resources or tools.

In addition, there exists prior work regarding initial best practices and patterns for building GenAI based applications. In~\cite{fowler2024genaipatterns}, several established lower-level design patterns are listed to improve and evaluate the quality of LLM-based applications, including fine tuning, RAG, better retrieval methods or LLM as a judge mechanisms. In addition, Amazon AWS proposed an initial set of best practices for building robust GenAI appliations~\cite{ladeiraTanke2024awsbedrockagents1, ladeiraTanke2024awsbedrockagents2}, providing guidance on how to design better agents, and advocating for comprehensive logging, observability and testing capabilities.

While these examples illustrate the benefits of integrating traditional SE approaches with GenAI-driven methods, or provide partial guidance for developing LLM-based or agentic applications, there is an urgent need for a more holistic approach to design and develop robust, adaptive, and efficient GenAI-enhanced systems. In this paper, we outline a vision for a GenAI-native system as an architectural paradigm that integrates the cognitive capabilities of GenAI with the established operational excellence, reliability, and dependability principles of traditional systems. We will build upon existing SE paradigms and GenAI best practices, indicate where they fall short, and propose several new principles and patterns. 

In this paper, we will not focus on recent coding paradigms~\cite{sapkota2025vibecodingvsagentic}, tools~\cite{github2021copilot, anthropic2024claudecode, cursor2024, continue2024}, frameworks~\cite{zhu2025adacoder, openai2025codex}, or platforms~\cite{wu2024devin, stackblitz2025boltenew} associated with LLM-based coding or SE assistance (e.g., LLM4SE) aimed at aiding or automating the development and maintenance of software artifacts~\cite{hou2024large, zhang2023survey, rao2025softwareengineeringlargelanguage}. Our primary focus is on exploring the impact of GenAI on future software design principles and patterns. However, in Section~\ref{sec:sw_stack}, we will briefly address how existing cloud-native and agentic cloud platforms could be further enhanced. 

\vspace{-2mm}
\section{Example Use Cases and Applications}
\label{sec:examples}

We claim that the vision and contributions outlined in this paper will be widely applicable across diverse use cases and application domains, spanning all layers of the stack and phases of the software development lifecycle. To better illustrate our vision and concepts, and to demonstrate the limitations of traditional or purely agent-based approaches, we first introduce a few selected example use cases. 

\subsubsection*{GenAI-native Micro-functions}

Imagine a GenAI-native contact information parsing function. Unlike traditional implementations that only accept limited sets of inputs or modalities based on predefined structural rules---such as those easily parsed through pattern matching---a GenAI-native implementation should efficiently and reliably accommodate a broader spectrum of inputs across modalities, including unstructured text, YAML, or images, without assuming fixed input formatting and structure. Furthermore, it should adeptly handle incomplete or inaccurate information. As discussed later, dependent or downstream functions should become resilient, anticipating potentially incomplete or uncertain parsed data, instead of relying only on the accuracy or certainty of outputs from the parsing function.

\vspace{-2mm}
\subsubsection*{GenAI-native Web Applications}

At the software system level, the future of GenAI-native web services and applications can be reimagined to offer greater flexibility and personalization. Unlike traditional web services, which typically provide specific, restricted, and rigid APIs and user interfaces for interacting with other internal or external services, a GenAI-native approach would enable both end users and other services to customize or personalize the interface and behavior dynamically. This customization could happen on-the-fly while maintaining the reliability, scalability, and safety properties inherent in traditional web services.

Such customizations can be either temporary or permanent and may necessitate bespoke communication, processing, storage, and user experience rendering capabilities. Note that this approach will also require adaptive specification of requirements, along with clear accountability and responsibility agreements with the end user or across such services.

Imagine a user of a task list management service wishing to integrate weather information into their task list entries, extending beyond the core functionality supported by the service. Or similarly, an external service requesting additional information or attributes from a weather service, exceeding the traditional API's capabilities. In both scenarios, the target service must determine whether to accommodate such enhanced capabilities, establish the conditions under which they will be supported, and undertake steps to implement these custom features or experiences. Some requests may be provided almost instantaneously as a bespoke request, whereas others may require more time and effort.

\vspace{-2mm}
\subsubsection*{GenAI-native Software Upgrades} 
Microservice software upgrades could also be transformed into a fully GenAI-native process, encompassing both the initiation and execution phases. Instead of relying on traditional human-centric performance evaluation and optimization loops, a GenAI-native service could autonomously monitor its interactions with other services to identify suboptimal usage of its functionalities or interfaces. In response, DevOps agents within the service may independently decide to develop a new service endpoint with improved core capabilities. After rigorous testing and integration, this enhanced functionality could then be deployed and proactively communicated to other services, allowing dependent GenAI-native services to seamlessly incorporate the new features into their core logic, either autonomously or with human oversight. This approach would facilitate more proactive and seamless upgrades, thereby improving overall system efficiency and adaptability.

\subsubsection*{Discovering Unknown Unknowns} 
Traditional anomaly detection techniques, for example in case of predictive maintenance, often focus primarily on \textit{known unknowns}. However, self-reliant GenAI systems may also help uncovering \textit{unknown unknowns}—previously undocumented issues. Such system would autonomously detect, explore, and verify potential issues, learning from past incidents to improve effectiveness. Confirmed anomalies should trigger appropriate bespoke actions, and be converted into \textit{known unknowns}, by automatically creating, testing, and integrating new anomaly detection routines using GenAI-native principles. 

\subsubsection*{Enhancing Legacy Services} 
When upgrading traditional applications to GenAI-native systems, a viable strategy would be to retain the proven reliable logic of the legacy system while enhancing it with GenAI-empowered functionality to increase resilience and adaptability to a wider range of usage scenarios beyond those achievable through traditional logic or rules. For instance, in a bank transfer service, the core logic should ideally remain untouched. However, GenAI-native enhancements could include AI agents to detect and respond to anomalies that are not easily identifiable through traditional heuristics, thereby providing enhanced security. A notable example might be detecting spurious transactions resulting from successful phishing attempts. These agents should learn and improve over time, adapting to new threats and safely reconfiguring decision-making processes as needed, with failsafe mechanisms to revert to legacy mode if issues arise.

\section{GenAI-Native Design Principles}

This section defines the guiding principles for designing and building GenAI-native systems. We begin by first defining the five foundational pillars onto which we will establish these principles. We draw inspiration from historical technology transformations and human organizational methodologies, and highlight key differences with traditional systems. 

\subsection{Design Goals}

\subsubsection*{Reliability} The capacity of a system to function correctly and predictably over time. It encompasses the ability to recover from (un)expected failures or disruptions (\textit{resilience}), handle unexpected inputs, conditions or stresses without failing (\textit{robustness}), and continue to function appropriately even when components fail or errors occur (\textit{fault tolerance}).

\subsubsection*{Excellence} The capacity of a system to achieve the highest standards in performance, quality, and effectiveness within a domain. It encompasses the ability to apply learned knowledge, skills and behaviors in a specific context or domain (\textit{competency}), produce consistent, predictable, and repeatable results (\textit{precision}), and execute processes optimally, and with minimal manual intervention, ensuring high-quality outcomes while maximizing resource utilization (\textit{proficiency}).

\subsubsection*{Evolvability} The capacity of a system to change, grow, and improve over time in response to internal or external factors. It encompasses the ability to adapt to new environments (\textit{adaptability}), incorporate incremental functional or structural changes (\textit{flexibility}), and undergo significant restructuring and redesign (\textit{malleability})~\cite{litt2023malleable}.

\subsubsection*{Self-reliance} The capacity of a system to handle things on its own and provide for itself. It encompasses the ability to solve its own problems and adapt to new challenges without heavily relying on external help (\textit{self-sufficiency}), autonomously perform tasks, make decisions, and operate independently without external control or continuous human intervention (\textit{self-governance}), and continuously self-enhance its performance and recover from issues through learning, optimization, or healing (\textit{self-improvement}).

\subsubsection*{Assurance} The capacity of a system to foster a secure and trustworthy environment in accordance with predefined standards. It encompasses the ability to address biases and ethical considerations (\textit{alignment}), protect against unsafe use, threats and vulnerabilities (\textit{security}), and maintain the trust, integrity and privacy of sensitive information and key stakeholders (\textit{trustworthiness}). 

\vspace{\baselineskip}

While many of these objectives are not unique to GenAI-based systems, the intrinsic characteristics and behavior of GenAI, coupled with the diverse range of tasks and inputs they can handle, make achieving these goals during the design, development and operationalization both critical and complex.
For instance, the probabilistic nature of GenAI-based solutions and their outcomes requires more comprehensive and elaborate reliability measures, both throughout development and during operational phases.
Additionally, self-reliant and evolving GenAI systems demand additional innovative assurance and performance measures to ensure safety and effectiveness.
Successfully realizing these goals requires several key innovations, including the development of new design principles, methodologies, techniques, and tools. 
In this paper, we advocate for a more systematic approach, rather than relying on ad hoc or manual strategies. 

\begin{figure*}[t]
  \centering
  \includegraphics[width=0.85\textwidth]{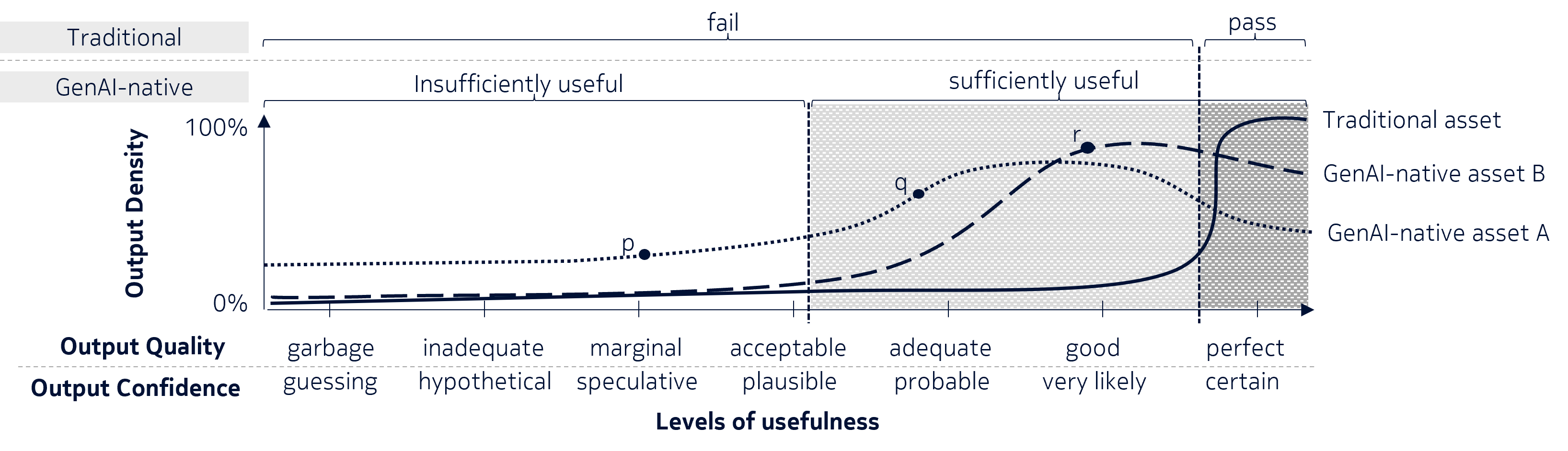}
  \caption{Example probability density functions of output quality or confidence for outputs produced by typical GenAI and traditional logic based solutions across the full range of acceptable inputs. }
  \label{fig:sufficiency}
\end{figure*}

It is important to emphasize, however, that not all aspects will be equally important across all application domains. Safety-critical AI domains may prioritize self-reliance, reliability, assurance and excellence, such as autonomous scheduling and planning systems, robotics, or industrial automation. Others may emphasize assurance, reliability and excellence, such as customer support chatbots, business intelligence, enterprise workflows. Meanwhile,  creative industries or recent AI developer tools may prioritize evolvability and excellence. 

\vspace{-2mm}
\subsection{Design Perspectives and Analogies}
\label{sec:analogies}
To better motivate these paradigms and patterns, we first briefly draw parallels with two historical transformations and reflect on how we organize ourselves to cope with imperfections while striving for excellence. 
First, when transitioning from circuit-switched to IP networking to enable more efficient use of network resources, higher-level mitigation strategies like forward error correction, packet reordering, and retransmissions were developed to retain reliable communication. These strategies compensate for issues such as corrupt or dropped packets and out-of-order arrivals.

Second, when shifting from stateful monolithic applications to cloud-native stateless microservice architectures to facilitate efficient utilization of hardware and system resources, mitigation strategies like cloud-native architectures, advanced lifecycle management, and new design principles were developed to address performance and latency issues. These strategies address the challenges introduced by best-effort virtualized resource access and time slicing.

Third, when drawing parallels to human organizations, enterprises and factories organize themselves to accomplish tasks effectively by promoting efficient utilization of human and automation capabilities. Efficiency and mitigation strategies like team collaboration, clear communication protocols, proper resource planning, regular training programs, and robust feedback mechanisms have been implemented to address potential quality and efficiency issues, caused by best-effort organizational, operational, and management methods, as well as inherent human error.

These perspectives underscore the importance of developing robust mitigation strategies and organizational mechanisms to ensure higher layers remain efficient and resilient against the inherent limitations and imperfections of the underlying technology. We firmly believe that a similar approach is essential for GenAI-native software and knowledge systems, to help overcome the inherent limitations of the technology and bypass the "generate and pray" strategy present in many existing GenAI solutions.

\section{GenAI-Native Best Practices}

In this section, we translate the high-level design goals and perspectives into a set of guidelines tailored for GenAI-native applications. Building upon established software engineering practices, we motivate why and how these practices should be extended or adapted to incorporate GenAI into future software design methodologies.  
While we do not claim that these guidelines are exhaustive, we hope they will serve as a robust foundation that will inspire the AI, programming, and software engineering communities to further refine and expand upon them.

\subsection{Reliability Guidelines}

We start with these guidelines, as we believe they are the most critical when designing and building GenAI-native systems, enabling a strong foundation for the other guidelines.

\subsubsection*{Design for Fault-tolerance and Resilience} 
Traditional software engineering solutions are often designed and developed according to clearly defined \textit{pass/fail criteria}, and are rigorously tested to ensure they reliably handle predefined input ranges without failure. However, this approach is unattainable for GenAI-native systems due to the nature of GenAI as well as the kind of inputs and tasks. 

Consequently, instead of relying on clearly defined pass/fail criteria, we propose the concept of \textit{utility-based sufficiency criteria}. This emphasizes the practical effectiveness and adequacy of a solution in real-world scenarios, reflecting solutions that are \textit{sufficiently useful most of the time} in terms of both output quality and inherent uncertainty.

A conceptual example is shown in Figure~\ref{fig:sufficiency}, where three probability density functions are depicted of the output quality and confidence likelihood across the entire set of acceptable inputs, for outputs produced via GenAI or traditional logic based solutions. Note that for similar tasks, the acceptable input range of traditional assets will typically be much narrower than for GenAI-empowered assets. 

Not only do samples $p$, $q$, and $r$ inherently have lower quality and/or confidence scores compared to the output from a traditional asset, sample $p$ would be considered insufficient due to subpar quality or confidence, while samples $q$ and $r$ would both meet the sufficiency criteria. In many cases, sample $r$ may be preferred, but in some cases, sample $q$ may be preferred due latency or resource utilization. 

Applied to the contact information parsing example, a traditional implementation will typically only accept a very restricted set of inputs, according to a predefined set of structural rules. Asset $A$ on the other hand could be a fully agentic implementation based on a simple AI model that can only reliably extract parts of the contact information. In contrast, asset $B$ could be GenAI-native implementation, capable of extracting more information with greater reliability and accuracy, possibly also with other runtime tradeoffs. 

This resembles also many professional human activities, where perfect outputs are not always attainable or necessary either. For example, in case of a human document summarization or detailed report creation task, different people will naturally produce different outputs, and we typically neither expect nor assume perfection. In many cases, this cannot even be uniquely measured.

Therefore, human produced outputs typically undergo a review process, according to task-specific evaluation criteria, which requires additional time and effort. 
GenAI-native systems should be architected accordingly.

\subsubsection*{Include Verification and Mitigation at all Levels} 
Given the inherent reliability challenges of GenAI assets, it is crucial for GenAI-native systems to integrate thorough verification and mitigation strategies throughout the software stack and software lifecycle. 
Effective strategies include the native integration of design time and runtime self-verification and fact-checking mechanisms, as well as the use of external verification systems and tools.

Even more than in traditional systems, dependent assets should not presume the reliability or predictability of syntactically or semantically correct outputs, even when a GenAI asset asserts the implementation of self-verification and self-mitigation strategies. 
This may require incorporating cognitive, probabilistic, or approximate capabilities into dependent assets to interpret and assess the usefulness of received results at runtime, as well as implementing additional quality absorption or adaptation strategies, which often may partly require them to be a GenAI-native assets as well. 

Mapping this onto our example use cases, a GenAI-native web service or parsing function, as well as its dependent services or functions should always anticipate potential communication, information or processing issues. In addition, they should implement appropriate sanity checks and mitigation strategies, not only through extensive pre-production testing, but also at runtime, and in production.

\subsubsection*{Restrict Scope of Unreliability} 
Since mitigation strategies may involve absorption or delegation, sources of unreliability may easily spread across assets at runtime. While this can be acceptable, for example when latency and throughput are prioritized over highly accurate outputs, it is advisable to restrict the scope of unreliability. 

Beyond traditional circuit breakers from distributed systems, a GenAI-native variant may first implement additional conversational mitigation strategies, including requesting the upstream asset to partly redo or improve computations, or falling back to more conservative approaches, before giving up on the asset altogether. Additional considerations include the ability to induce and measure mutual progress, assess convergence, and possibly consider game theoretical strategies such as the Nash equilibrium~\cite{nash1951non}.  

This approach mirrors how humans typically collaborate within teams or functional units, where imperfection and multiple iterations are generally well accepted internally but less so across teams. 
In case of GenAI-native web services, web services should be designed with proper strategies to cope with inherently unreliable dependent services, not only because of networking or stability issues, but also because of unreliable semantic communication and processing.

\subsubsection*{Promote Transparency of Processing and Risks} 
Instead of only returning the results with other assets through traditional interfaces, GenAI-native assets should be more transparent when sharing their results with other assets, allowing the latter assets to more easily assess the usefulness of the provided results.  
Transparency may include sharing the applied processing paradigm (e.g., AI-based versus traditional), providing an interpretation of the request and explaining the execution or reasoning process, as well as sharing self-verification and mitigation strategies used during its processing. 
Assets should negotiate the amount of risk, transparency as well as responsibility and accountability before interacting. In additional they should (proactively) send additional metadata alongside the actual output, or interact through conversational paradigms.

In our example use cases, based on additional provided metadata, downstream assets can evaluate the reliability and quality of parsed contact information, or assess the usefulness of answers provided by GenAI-native web services. This evaluation may include determining whether the task was completed using traditional or cognitive processing, evaluating the provided confidence score, or reviewing any additional execution comments or requests for further clarification made by the asset.

\begin{figure*}[t]
  \centering
  \includegraphics[width=0.90\textwidth]{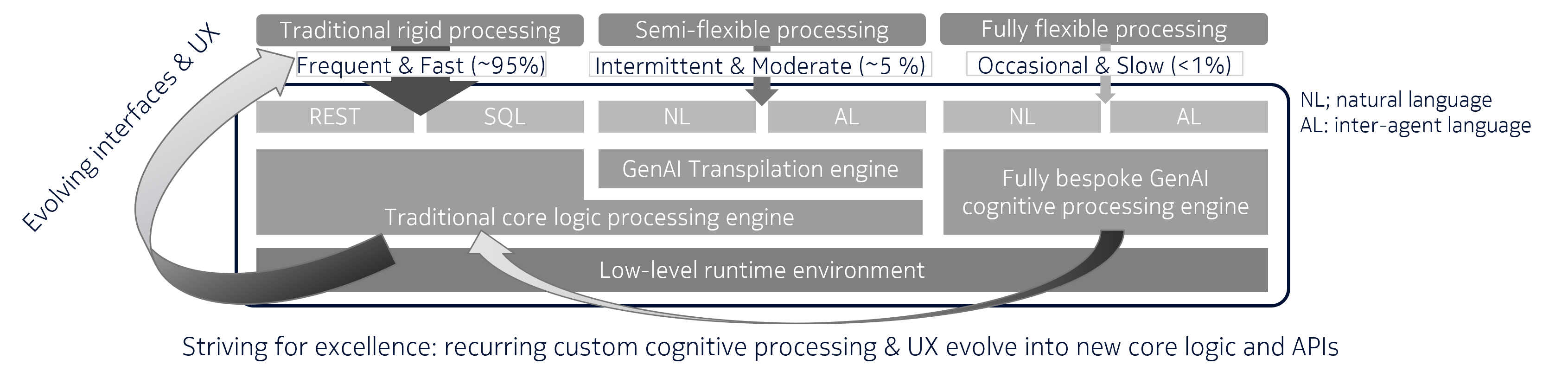}
  \caption{Conceptual view of a self-improving hybrid GenAI-native asset: fast routine traditional processing \& interfacing, and slow occasional (semi-)cognitive processing (cfr. System 1 and System 2 thinking~\cite{kahneman2011thinking}), with gradual optimization loops.}
  \label{fig:excellence}
\end{figure*}

\subsubsection*{Plan for Contingencies} 
Similar to distributed systems, GenAI-native assets and subsystems should integrate contingency strategies to address unexpected issues or errors. 
One example strategy includes incrementally generating several outputs using diverse techniques, resources or assets. As a result, it is essential to plan for adequate processing margins and slack time to improve overall outcomes or recover from anticipated challenges. 

For instance, downstream assets might opt for more costly or slower methods to accurately extract specific contact information or to perform some web service. Unlike distributed systems, the challenges in GenAI-native systems are more likely to arise from interaction and processing complexities rather than networking or performance-related issues.

Like human contingency methodologies, assets should plan ahead, preparing and executing tasks well before the deadline, considering the estimated time and resources required for potential refinements.
As discussed later, we implicitly assume that GenAI-native assets and services will increasingly operate asynchronously, proactively, and in parallel, akin to a true distributed or federated system, rather than traditional reactive micro-service architectures.

\subsubsection*{Minimize Dependency on Cognitive Processing}  
We believe that, in general, a foundational guideline is to minimize dependency on cognitive processing within all assets, either from agents or human-in-the-loop, effectively reducing the level of \textit{artisanal} processing. This approach, illustrated in Figure~\ref{fig:excellence}, contrasts with a current trend of GenAI-first agentic solutions, where agents are central to most critical processing paths. Example strategies include systematically identifying and writing dedicated code snippets, or training a more narrowly scoped yet efficient AI/ML model to replace common cognitive workflows. 
In our example use cases, dependency on (generic) cognitive processing should be systematically reduced by identifying which inputs or service requests can be handled by more efficient and reliable processing methods and interfaces.

Reducing open-ended cognitive processing not only serves as an excellent mitigation strategy for enhancing reliability, it can also improve operational excellence, as well as mitigating assurance issues. 
Finding the optimal balance between cognitive and traditional processing will require the development of additional domain-specific and task-specific expertise, tools, and techniques. 
We can draw inspiration from how we discover and organize ourselves to automate and harden repetitive tasks, while specific human processing can augment these solutions to provide bespoke outcomes or customize existing repetitive tasks.

While exceptions to this guideline exist, such as in search or optimization tasks where large language models (LLMs) or reasoning models can expedite processes compared to traditional methods, it remains crucial to determine the optimal balance between GenAI and traditional assets.


\subsection{Excellence Guidelines}

In addition to enhancing the reliability, it is crucial to also consider quality and efficiency—two often conflicting requirements—during design and operations. The following principles provide guidance on balancing these requirements. 

\subsubsection*{Build upon Proven Design Principles and Practices} 

While it may seem evident, it is crucial to emphasize the importance of leveraging established software engineering and organizational methodologies and practices to systematically improve quality and efficiency. For instance, process methodologies such as incorporating checklists, continuous software testing and CI/CD pipelines, or standard operating procedures (SOPs) help ensure critical processing steps are not overlooked and prevent reinventing the wheel. GenAI-native assets should not only build upon these through proper tools and frameworks, they should also be revised, optimized and extended to allow seamless blending of traditional and cognitive processing. 

In our example use cases, it could be beneficial to implement checklists or customized validation procedures to verify the accuracy of the generated results, such as the parsed contact information or the return messages from GenAI-native web services. More importantly, since GenAI models have a limited lifespan, swapping the model underneath GenAI assets or multi-agent systems will inevitably alter their \textit{personality}—that is, their behavior and capabilities. This can lead to ripple effects, including potential stability and convergence issues. As it is unlikely that such updates will produce a service identical to its predecessor, robust design, testing and service evolution principles will be essential to prevent significant disruptions.

\subsubsection*{Optimizing Cognitive Workflows} 

Systematically reducing cognitive processing can significantly enhance efficiency and reliability, leading to more effective and streamlined operations. This involves transforming repetitive or time-consuming cognitive tasks, eliminating unnecessary steps, and preventing recurring chain-of-thought reasoning. Traditional workflows can substantially improve runtime efficiency while also minimizing the need for additional costly reliability measures. Furthermore, this approach may mitigate the impact of swapping the underlying GenAI model in a GenAI service, as most repetitive tasks implemented through traditional methods would remain unaffected. However, this will necessitate adaptive routing and handling capabilities across all request types and modes.

Applied to our example use cases, traditional parsing methods, service handling and API protocols should be leveraged as much as possible to efficiently and reliably communicate and handle known input requests.  Ideally, this process is fully dynamic and adaptive, rather than being hardwired. 
Specifically, in the context of traditional APIs, MCP, and A2A protocols, assets should ideally prioritize interaction through traditional APIs for both communication and processing. This approach ensures reliability and consistency. In addition, assets could also serve as dependable MCP resources or tools towards other GenAI assets. Full A2A communication and processing on the other hand should be reserved when it is strictly necessary.

\subsubsection*{Systematic Quality Verification and Retrospectives} 
In addition to optimizing cognitive workflows, systematic quality review, verification, and improvement can significantly enhance quality and efficiency~\cite{ladeiraTanke2024awsbedrockagents2}. This may be achieved by incorporating continuous learning and feedback loops at strategic points, including at runtime, leveraging and rethinking established methodologies such as Kaizen~\cite{imai1986kaizen} and Six Sigma~\cite{pyzdek2023sixsigma}, performing regular code quality and maintainability checks, and allocating dedicated time blocks for targeted experimentation.

In our example use cases, it is crucial to consistently evaluate the quality, efficiency, and effectiveness of GenAI-native functions and services. This involves assessing the performance of GenAI assets against new input types, or monitoring their evolving capabilities. Consequently, GenAI-native systems will demand more comprehensive and sophisticated verification procedures than traditional software systems, often dependent on advanced cognitive methods. Striking the right balance between these approaches will be essential for optimal performance and effectiveness.

\subsection{Evolvability Guidelines}

A unique capability of GenAI is its ability to generate innovative solutions for new problems or tasks. This enables GenAI assets to be flexible and adaptive, creating new solutions and improving existing ones. However, this also requires a sufficient degree of restraint and discipline to avoid chaos. 

\subsubsection*{Promote Resilient and Adaptive Designs} 
Instead of returning an error codes or throwing exceptions, as is common in traditional systems, GenAI-native assets can be designed to be more resilient and adaptive to unexpected requests and conditions. For instance, if a dependent function or service is malfunctioning, overly restrictive, or unavailable, the asset should leverage its cognitive processing capabilities to actively resolve the issue and find a solution, rather than simply throwing an exception. 
Similarly, in case of GenAI model swaps or other behavioral changes, dependent assets should be designed to be tolerant and adaptive against such changes.

In our example use cases, GenAI-native functions or services should be designed to (temporarily) adopt alternative processing or communication strategies when encountering issues. 
In the worst-case scenario, they should be able to switch to another function or service to bypass the problem. For instance, if a web service relies on a weather or location service that becomes unresponsive or cannot handle a custom request, or its \textit{personality} has altered, the service should either adjust its strategy or seek an alternative service capable of fulfilling the request.

\subsubsection*{Evolve towards Reliable and Efficient Systems} 
Whenever resilient and adaptive cognitive processing and designs are required, the system should monitor and record these bespoke behaviors and interactions. If such behaviors occur frequently, or the cognitive processing becomes too expensive, the system should evolve its internal processing to natively support these capabilities, resulting in a more reliable, repeatable and efficient solution. In other words, a GenAI-native system or asset should learn to evolve in a manner that systematically reduces its dependency on cognitive processing and bespoke solutions, as shown in Figure~\ref{fig:excellence}. For our example use cases, this could mean gradually expanding the core functionality of the parser function or web service.

\subsubsection*{Promote Consistency over Creativity} 
Unless required, GenAI-native systems should restrain their creativity, bespoke strategies, and solutions. Though obvious in traditional systems, consistent, repeatable and dependable behavior should always be preferred over overly creative agentic solutions, especially for common tasks and scenarios. Such approach not only will enhance system stability but also reduces inconsistencies between assets, such as constantly needing to compensate for custom cognitive interactions and responses. This will also result in a smoother experience, for example in case of evolving web frontend designs, as people generally prefer stable and predictable look-and-feel and behavior. 

Applied to our use cases, the evolvability of GenAI-native micro-functions or web services should be limited, emphasizing consistency and resilience over creativity. Essentially, these services and functions should maintain their specialization, rather than evolving into generalist agentic solutions. Note that this approach does not preclude a GenAI-native system from incorporating a variety of generalist services as well, similar to human organizations.

\subsubsection*{Collective Competency Ecosystems} 
A potential key advantage of AI agents over humans is the efficient transfer of new competencies (e.g., knowledge, expertise, or skills). Instead of having assets independently rediscover similar competencies, a GenAI-native system should facilitate easy, efficient, and safe sharing of new competencies within and across systems. This allows individual assets to explore new competencies while benefiting from collective advancements. 

In our parsing example, more reliable or efficient contact information parsing prompts, methods or extensions for particular inputs could be shared with other systems to avoid rediscovery. This may in a further evolution and convergence of existing open source and marketplace ecosystems, where constantly evolving assets, beyond standalone model hubs or code repositories such as HuggingFace or Github, can be actively shared and ingested, respectively.


\subsection{Self-Reliance Guidelines}

A key objective of GenAI is to develop fully autonomous systems that reduce the reliance on restrictive, hand-crafted solutions. Given the transformative potential of such future systems, it is crucial to exercise caution and restraint. Although the discussed guidelines are not unique to GenAI-native systems, it will be essential for deeply integrating them into the code design and operations of such systems.

\subsubsection*{Balance Autonomy with Safety and Control} 
Self-reliant GenAI-native systems must implement clear decision-making frameworks and policies to govern all autonomous actions. 
Ideally, they should be made aware of such policies to allow for anticipation instead of continuously encountering opaque barriers. This requires additional safety checks and validations to detect ad\-ver\-sa\-rial attempts to circumvent the policies. 
Furthermore, internally and externally triggered measures should be implemented in case of repetitive misbehavior, such as penalizing, replacing, or disabling the asset. 
In our example, this may limit the ability of the GenAI-native contact parsing function or web service to evolve or even restrict its usage if it becomes too unreliable.

\subsubsection*{Include Rollback Capabilities for Autonomous Actions} 
GenAI-native assets should be designed with robust rollback mechanisms to address the impact of suboptimal autonomous decisions and evolutions. This includes the ability to revert or unlearn newly acquired competencies, actions, interfaces, or data representations. For instance, it should be straightforward to undo ineffective APIs or processing enhancements made to web services.

\subsubsection*{Perform Regular Reviews and Updates} 
All autonomous or self-reliant agent behaviors should be regularly reviewed internally and externally to detect potential issues that may have gone unnoticed. Based on these reviews, policies may be updated, and assets may be required to adjust or partly revert their behavior accordingly. Ideally, these processes should include mechanisms to learn from past incidents and actions, preventing endless loops of recurring violations.

\subsubsection*{Maintain Visibility into Autonomous Operations} 
To facilitate review and rollback in autonomous systems, log trails of all actions and decisions, including self-diagnostics and monitoring at multiple levels, and possibly involving human oversight, are essential. For instance, if an asset decides to rewrite parts of its core logic, or wants to access new services, this should be logged and potentially gated. Human organizations maintain clear rules of engagement regarding the levels of autonomy and self-reliance that are permitted versus those that require prior approval. 


\subsection{Assurance Guidelines}

Implementing required assurances within GenAI-native systems is crucial and impacts all other pillars. 
This section highlights key guidelines, recognizing that further effort and consideration is required to fully understand all aspects.
Though these are obviously also not unique for GenAI-native systems, GenAI imposes several new and hard thread vectors that need to be dealt with accordingly.

\subsubsection*{Adopt GenAI Security Best Practices} 
The OWASP guidelines~\cite{owasp2025genai} highlight several threats and issues related to LLMs and agentic processing. Example concerns include insecure output handling, supply chain vulnerabilities, excessive agency, and sensitive information disclosure. It is prudent to assume that all GenAI-native assets may be internally or externally contaminated or compromised, whether deliberately or inadvertently. As overreliance on cognitive processing can easily exacerbate the problem, this guideline underpins the main rationale of this paper: to carefully balance traditional and cognitive processing, and to handle all remaining cognitive processing with the utmost scrutiny.

\subsubsection*{Provide Observable and Explainable Solutions} 
GenAI assets should implement sufficient observability, interpretability, and explainability mechanisms. Examples include introspectability of cognitive techniques, as well as readability of all generated code, reasoning plans, and prompts. These mechanisms should be actionable, allowing for immediate remediation as well as long-term maintainability, reviews, and optimizations.

\subsubsection*{Design for Privacy, Integrity and Trust} 
In addition to enforcing security best practices and ensuring transparency, GenAI-native assets that process or exchange sensitive or confidential information must be tightly managed and controlled. Practical solutions include providing appropriate guardrails, running these assets within secure and well-guarded sandboxes, and tightly manage and restrict the information that can be processed within these sandboxes, or that can flow outside of them.

\subsubsection*{Assess and Mitigate Misalignment} 
As cognitive processing is highly susceptible to biases, proactive and reactive measures are essential to prevent, detect, and hopefully mitigate these issues. Measures include carefully selecting the underlying GenAI technologies, incorporating active guardrails and filters, and regularly conducting focused audits on outputs, intermediate reasoning, processing, and generated code.


\subsection{Architectural and Operational Guidelines}
\label{sec:arch}

In this section, we outline key architectural, organizational and operational guidelines for designing systems that adhere to the core GenAI-native guidelines discussed earlier. 

\subsubsection*{Reimagine Cloud-native Paradigms} 
While cloud-native design principles and patterns provide a solid foundation for designing, developing, and managing GenAI-native systems, they need to be rethought to accommodate the unique characteristics of GenAI-native systems. For instance, cloud-native principles lack the flexibility to efficiently handle the inherent malleability, reliability, and self-reliance aspects of future GenAI-native systems. Concepts such as immutable infrastructures, rigid service meshes, and fixed API-driven service interfaces, though useful as foundational principles, need to be generalized, relaxed, or extended. 

Specifically, immutable infrastructures should be evolve into \textit{reproducible organic infrastructures}, where GenAI-native assets can freely evolve at runtime, while retaining easy and efficient replicability in case an asset fails or needs to be scaled. Similarly, service meshes should evolve into \textit{organic substrates}, enabling assets to interact more freely while remaining compliant. Finally, fixed API-driven service interfaces should transform into \textit{unified conversational interfaces} (UCI), comprising an organic mix of traditional yet dynamic APIs, as well as conversational communication mechanisms. 

\subsubsection*{Create Future-proof Designs} 
GenAI-native designs should reflect the organic and evolvable nature of GenAI. This involves transforming existing micro-service designs into more modular micro-function designs, to facilitate independent evolution of core functionalities and interfaces, but also creating future-proof internal data representation schemas. This also involves anticipating changes in the constellation and communication patterns among assets, akin to the organic evolution of human teams and their interactions. Future-proof designs should also allow for easy rollback of code, state, as well as data representations to earlier versions.

\subsubsection*{Relax Strict Application and System Boundaries} 
Similarly to loosely coupled distributed systems, strict boundaries between GenAI-native assets and systems should be relaxed into a more organic substrate or organizational structure. Through evolvability and self-reliance, and driven by a continuous pursuit of excellence, GenAI-native systems should allow for dynamically adapting its organisational structure, such as automatically switching to alternative services with better capabilities or behavior, integrating novel services to accommodate custom requests, and more. 

GenAI assets allow for more easily interchangeable communication protocols (cfr. MCP or A2A), enabling GenAI-native (web) services to more easily swap out one service for another, even though the alternative service may have a very different personality providing somewhat different functionality. For example, if a user wishes to incorporate weather information into a GenAI-native calendar service, the system should autonomously determine how to contact a weather service and integrate this functionality seamlessly.

\subsubsection*{Impose Clear Scope and Responsibilities} 
Building a robust GenAI-native system will require strong organizational principles to ensure stable behavior. This requires clear governance to establish the scope and responsibilities for all assets, their interactions, and the extent to which they are allowed to evolve. This requires actively monitoring all GenAI-native assets as well as imposing restrictions.  

For example, if an asset autonomously decides to communicate with a non-approved service, makes unauthorized changes to core functionality, uses non-interpretable interfaces and protocols, or is clearly exceeding its intended scope or responsibilities, the system should intervene and compel the asset to take corrective actions. This may include shutting down the asset or forcefully reverting it to an earlier state. Traditional organizational software engineering patterns may be extended with human organizational structures to better orchestrate and coordinate these aspects.

\subsubsection*{Model GenAI Agents as Digital Workers} 
In AI-first multi-agentic designs, agents are often positioned at the center of the system, where all decisions and processing flow through them, and the actual resources are accessed via tools.  
In our vision for GenAI-native systems, agents should instead adopt a collaborative worker role, focusing on the resources they manage and operate. These agents should continuously strive to optimize their involvement, minimizing their presence in any critical path unless necessary. Much like a bee colony working together towards a common objective, these agents should facilitate seamless collaboration and efficiency.

\begin{figure*}[t]
  \centering
  \includegraphics[width=0.8\textwidth]{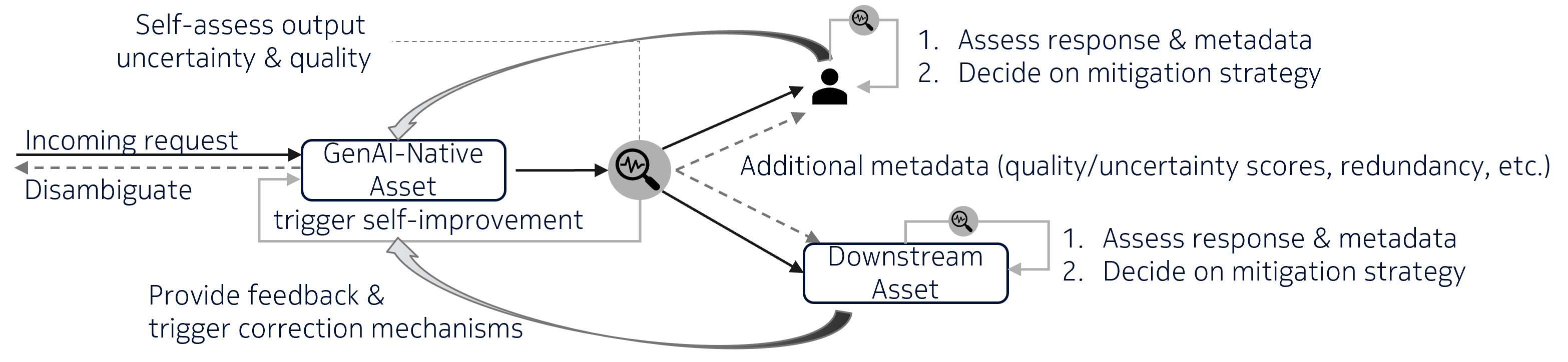}
  \caption{GenAI-native blueprint comprising several patterns for enhancing resilience to quality and uncertainty variations. }
  \label{fig:embrace_uncertainty}
  \vspace{-3mm}
\end{figure*}

\subsubsection*{Create Self-contained GenAI-native Capsules} 
As will be discussed in Section~\ref{sec:design_patterns}, a GenAI-native asset comprises multiple components, both active and passive, akin to a biological cell. Proper orchestration and management of these assets requires an evolution of existing cloud-native patterns. 

\subsubsection*{Adopt Organic Lifecycle Management} 
The lifecycle management of GenAI-native systems should natively incorporate evolvability and self-reliance. Next to traditional, planned offline application enhancements and optimizations, it is expected that the traditional boundaries between software development and CI/CD cycles will blur, as GenAI-native assets will be able to modify actively deployed assets. To better support this, DevOps and XOps practices should be enhanced to empower hybrid human and agentic application codesign. CI/CD approaches should become bidirectional, allowing online GenAI-native assets to actively contribute to the system, rather than merely receiving pushed updates.


\subsection{Programmability Guidelines}

Developing adaptive, self-evolving GenAI systems requires specialized programming practices across the full stack.

\subsubsection*{Develop Organic Programming Paradigms} 
GenAI agents should be able to easily make small, targeted changes to an existing code base, application interfaces or data structures, even while the application is running. Therefore, we envision a more modular micro-function programming paradigm, with tiny functional and data representation constructs that have a clearly defined purpose, and can easily evolve or be rolled back. Such paradigm should also allow human and agentic developers to easily distinguish between core and organic functionality, including separating the traditional from the cognitive parts. Additionally, innovative organic programming techniques must be developed to better cope with aspects like \textit{sufficiently useful}, probabilistic and evolvable outputs produced by other assets.

\subsubsection*{Develop a Flexible Policy Language} 
A crucial aspect to successfully manage and control GenAI-native systems, is to be able to easily yet flexibly define and constrain the degree and scope of malleability and self-reliance of GenAI-native assets. This capability needs to be available at all levels, should be easy to express, customize (possibly autonomously), and enforce. Assets themselves should be aware of their capabilities and restrictions, possibly through machine readable contracts, protocols or agreements. Some of these may be the result of active negotiations across GenAI-native assets, whereas others may be the result from external organizational entities enforcing these onto these assets.

\section{GenAI-Native Design Patterns}
\label{sec:design_patterns}

In this section, we present a set of initial behavioral, structural and creational design patterns based on the design principles and guidelines described earlier. We first focus on lower-level enabling patterns before integrating them into higher-level architectural and operational patterns. 

Although several of the proposed patterns are not exclusive to GenAI-native systems, as we will discuss, these patterns will often be more impactful, intrusive, and vital for maintaining a robust and reliable evolvable GenAI-native system, especially considering the inherent nature of the GenAI technology and the intended future applications.


\subsection{Reliability Patterns}

\subsubsection*{Reflective Processor} 
This pattern advocates for the inclusion of meta-cognition and self-regulation mechanisms, and to take necessary corrective or mitigative actions rather than blindly accepting them. This may involve disambiguating the task before actual processing occurs. This may also involve triggering additional pre and postprocessing steps, retrieving additional information or more context, running additional verification, involving other assets, or simply absorbing the reduced quality or confidence. It is crucial to recognize that such reflective processing may also be necessary for dependent non-GenAI assets.

\subsubsection*{Reflective Communicator} 
This pattern involves creating transparent and redundant bidirectional communication channels between assets. Transparency measures may include transmitting self-assessment scores or reports as additional metadata, either by the sending asset or as feedback from the receiving asset. Redundancy measures may involve producing multiple outputs, accompanied by additional context information, rather than a single output.
Both patterns are illustrated in Figure~\ref{fig:embrace_uncertainty}. Combined, they form the foundation of a robust GenAI-native system.

\subsubsection*{Resilience Fender} 
Aside from the traditional circuit breaker pattern from distributed systems, GenAI-native assets should also account for (accumulated) expected uncertainty or quality degradation issues, and implement mechanisms to either absorb or mitigate issues. This may include actively forcing upstream assets to partially or completely redo some of the processing. GenAI resilience fenders may range from being very flexible, in case the asset is able to absorb larger issues, towards being extremely firm, for example when integrating with legacy non-GenAI assets and services, where reflective processing and communication may not be supported. They may also act as cognitive circuit breakers, for example to avoid problematic assets from bringing down the system.


\subsection{Excellence Patterns}

\subsubsection*{Programmable Router} 
We propose a flexible integration of traditional and cognitive processing, where traditional core logic handles routine workflows while cognitive processing manages more complex and exceptional cases. This approach is reminiscent of the \textit{thinking fast and slow} paradigm~\cite{kahneman2011thinking}. A key enabling pattern to achieve this is the inclusion of a highly efficient programmable router or switchboard. This router can be deployed early in the processing pipeline to determine the optimal handling method for each request. Alternatively, it can be configured to reroute internal processing, such as adaptively switching between fast and slow paths, to triggering additional processing or handle exceptional cases if needed. 

Such application-specific router must be highly efficient to preserve the benefits of traditional core logic, yet highly programmable to allow for continuous evolution.  This may involve an iterative or hierarchical decision process, ranging from directly invoking core logic via specific APIs, over efficient LLM based domain and task routers~\cite{tran2025arch}, to agents reasoning about how to decompose and resolve a request. In each case, the router should be able to select between core logic functions or agentic implementations, and have the option to reroute when needed.

\subsubsection*{Continual Self-reflection} 
To ensure and enhance quality and efficiency, GenAI-native assets should implement several quality assurance mechanisms. These may include incorporating feedback loops, enabling and executing auditing trails, or running self-consistency checks. The results of such measures can subsequently trigger one or more evolvability or self-reliance patterns. This pattern is comprised in Figure~\ref{fig:embrace_uncertainty}.


\subsection{Evolvability Patterns}

\subsubsection*{Unified Conversational Interface} 
Instead of only relying on rigid APIs or natural language (NL) interfaces, GenAI-native assets should adopt a hybrid approach,  
where communication may happen via NL or via dedicated protocols both assets agree upon. This enables a more flexible interaction, beyond what is available through rigid APIs.
 
A core feature of such unified interface is the ability to gradually evolve from freeform conversations towards more traditional APIs. This implies that the latter API should not remain static but will gradually evolve over time. For efficiency and reliability, GenAI assets should encourage the reuse of existing APIs and logic, rather than continuously developing custom interfaces and logic. 

\subsubsection*{Cognitive Workflow Optimizer} 
This pattern involves the systematic identification, formalization, and transformation of key cognitive communication and processing patterns into traditional workflows. When integrated into the end-to-end functionality, it is also necessary to reconfigure the programmable router to correctly route relevant requests to these new workflows.   
To prevent the proliferation of highly specific and difficult-to-maintain workflows, it is essential to systematically review and refactor such workflows.

\subsubsection*{Organic Service Broker}
In traditional software systems, dependent assets are often tightly integrated, resulting in strong hidden dependencies. When a dependent assets fails, is unavailable, or cannot handle the request, the depending asset typically only can raise an exception.
This pattern allows for easily and dynamically switching to alternative functions or services, either to improve or to ensure continuous yet possibly degraded functionality. 
This involves detecting when to switch and finding other good alternative services or functions. This will typically also require additional cognitive processing to help retrofitting the new dependency and adapt subsequent processing and communication.

\subsubsection*{Malleable Data} 
This meta-pattern advocates for the creation, storage, and management of adaptable and evolvable data structures, representations, and handling methods to effectively address dynamic and evolving requirements driven by the systematic evolution of GenAI-native service logic, inputs, and modalities. Concrete patterns, paradigms and solutions will need to ensure that the evolution of such data organization and modeling incurs minimal disruption to the core logic. Furthermore, to enable seamless updates and facilitate easy rollback, solutions should support extensive versioning.


\subsection{Self-Reliance Patterns}


In this section, we introduce several high-level patterns for effectively managing self-reliant assets, focusing primarily on ensuring their reliability, security, and continuous improvement. While these patterns are not unique to GenAI-native systems, existing frameworks and mechanisms must be extended and fortified to address the unique challenges posed by GenAI.

\subsubsection*{Asset Lifecycle Management Teams}
Like in a human software company, multiple agentic or hybrid human/agent roles and teams will be responsible for managing the entire lifecycle of a GenAI asset. Example teams include asset management teams, asset development teams, and customer support teams. Some of these roles or teams may operate externally to the asset, while others may be, at least partially, directly integrated into the production asset itself.

For instance, when the asset receives a custom request that necessitates unforeseen capabilities, effort, or support, a team comprising agents and/or humans may be deployed to evaluate the short-term or long-term impact on the system. These teams may opt to temporarily allocate additional resources or initiate bespoke workflows to address the request. Alternatively, teams might choose to invest resources in expanding or enhancing the core asset's capabilities, following a thorough cost-benefit analysis and alignment with internal policies. For simple bespoke requests with minimal expected impact or overhead, the system may decide to skip many of these steps or use a very lightweight procedure.

\subsubsection*{Behavioral Policy-driven Safeguards}
Self-reliant assets should be explicitly informed of their permissible actions and restrictions through a set of static and dynamic guidelines, rules, or feedback. By sharing these rules of engagement through specific channels, assets can proactively learn to operate within these boundaries and automatically adapt to any changes. In addition to these rules, robust guardrails and controls must be provided to detect and prevent unintended, malicious, or wasteful behavior. This integrated approach ensures compliance, enhances operational efficiency, and maintains the integrity and security of the assets, providing a comprehensive framework for reliable and secure asset management.

For example, in case an external web service has a very specific request that would require extensive additional effort or trigger unforeseen actions, the system should be able to recognize such requests and act according to provided policies, such as denying such request or enforcing human oversight. Similarly, in case a self-reliant asset or agent wants to perform particular sensitive actions, appropriate safeguards should automatically kick in, for example by gate keeping access to specific systems, tools or other resources.

\subsubsection*{Infallible Fail-safe and Recovery Mechanisms}
Future self-reliant systems should incorporate inescapable fail-safe mechanisms to abruptly halt any further autonomous behavior in the event of abnormal, erratic, inappropriate, or highly inefficient actions. This is crucial to prevent further damage. In addition to a fool-proof built-in emergency shutdown functionality, various mitigation or recovery scenarios should be automatically triggered to maintain normal operation. Mechanisms such as rolling back to an earlier version of the asset, switching to a more conservative solution, or reverting to a fail-safe mode that provides minimal functionality ensure continuity and stability, safeguarding the system against potential disruptions.

\subsubsection*{Comprehensive Logging and Introspectability}

Assets must maintain detailed logs of all self-reliant actions to facilitate visibility, auditing, and retrospective analysis of the efficacy of such behavior. These logs should document all actions taken, their rationale, effort involved, and impact. Additionally, self-reliant assets should expose themselves to external inspectability by authorized entities through dedicated communication channels or direct inspection and diagnostic capabilities. This framework enhances transparency, accountability, and trust, allowing insights gained from analyses to trigger improvements and ensure continuous enhancement of the asset's performance.

\begin{figure*}[t]
  \centering
  \includegraphics[width=0.8\textwidth]{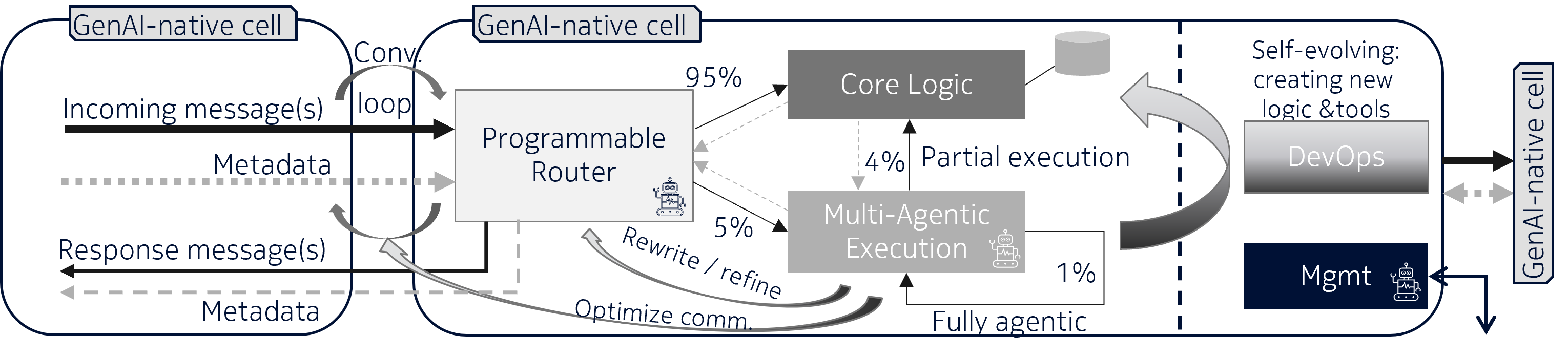}
  \caption{This figure illustrates the inner workings of a GenAI-native service, knowledge, or cyberphysical cell.  }
  \label{fig:cell}
\end{figure*}


\subsection{Assurance Patterns}

\subsubsection*{Cognitive Screening} 
Outputs originating directly or indirectly from GenAI-native assets should undergo thorough quality, security and safety screening before being accepted for further processing. This requires implementing robust encapsulation and gating mechanisms, spanning across all layers.  
Depending on the mutual trust, reliability, and sensitivity of the output or task, adaptive screening mechanisms can be employed. For self-reliant assets, cognitive screening complements higher-level behavioral safeguards.

\subsubsection*{Cognitive Firewall} 
To prevent undesirable cognitive communications and interactions across GenAI-native systems, and to provide scalable oversight and compliance, these systems must implement cognitive firewalling mechanisms. These mechanisms should be layered on top of traditional micro-service firewalling to control access within service meshes effectively. Cognitive firewalling can be implemented via rule-based or cognition-based deep communication inspection mechanisms, as an analogy to deep packet inspection. As with cognitive screening, different and adaptive levels of scrutiny can be implemented across different GenAI-native assets, based on external policies. The sidecar pattern, known from service meshes, can be repurposed for implementing such cognitive firewalls.

\subsubsection*{Agent Sandbox}
Sandboxing encapsulates the runtime and processing of an asset within a self-contained environment, allowing strict control over the impact of GenAI-native assets~\cite{david2025sandbox}. One type of sandbox prevents adverse effects from malicious, misbehaving, or erroneous assets. Another type facilitates safe experimentation and analysis, possibly using a digital twin of the real environment. A third type, possibly implemented via trusted execution environments (TEEs), allows untrusted assets to operate on sensitive data by regulating information exposure, which is beneficial for federated systems and external auditing.

\subsubsection*{Cybersecurity and Compliance Units} 
GenAI-native systems, similar to human enterprises, should integrate internal cybersecurity and compliance units alongside external auditing to ensure security and scalable oversight. These units can be managed by both autonomous agents and human oversight, governed by transparent and easily programmable policy frameworks. Key components include cognitive firewalling, screening, reporting, and inspection. 


\subsection{Architectural and Operational Patterns}

\subsubsection*{GenAI-native Cell}
\label{sec:cell}
We propose a key architectural and organizational pattern for designing and implementing GenAI-native assets, termed the \textit{GenAI-native cell}, as depicted in Figure~\ref{fig:cell}. The GenAI-native cell is a core building block that encapsulates many of the lower-level design patterns discussed earlier, encompassing all five foundational pillars. It represents a self-contained functional unit with a clear intent, to design and develop assets that are tolerant, adaptive, and strive for excellence.

Architecturally, a GenAI-native cell is an evolution of a microservice, and could be implemented as a set of sidecar containers, wrapped into one or more dynamic and adaptive service pods. This concept is partly inspired by a biological cell, consisting of a core (i.e., the nucleus), a semi-fluid cytoplasm containing various organelles supporting dynamic processes, and a cell membrane that interacts with and adapts to its environment.

Similarly, in the GenAI-native cell, we envision a static core, multiple dynamic processes, and adaptive interaction with neighboring cells. The core of a GenAI-native cell comprises all common logic, including traditional logic and AI/ML modules tailored to efficiently support specific subtasks or workflows. Alongside the core, we envision additional active and passive components to dynamically extend its capabilities. This includes one or more cognitive processing assets.

To interact with other cells, we envision an adaptive router responsible for handling and routing incoming and outgoing requests, and providing seamless handover between core and agentic processing, while considering reliability and resilience measures. To enable cells to evolve in functionality, processing, and communication efficiency, we envision active DevOps agents, whereas management components are responsible for managing and controlling the overall operations of the cells. Both DevOps and management assets can be deployed partly or completely inside or outside each cell.

\subsubsection*{Organic Substrate}

To flexibly configure, manage and control multiple GenAI-native cells, their evolution and their interconnectivity, we propose the concept of an \textit{organic substrate}, which represents an evolution of a service mesh or service chain. Within such substrate, cells can enter, exit, and possibly move across substrates over time. Logical or functional clusters of cells can be grouped and managed together as tissues or organs, each with higher-level joint purposes and goals, akin to human organizational structures.

The substrate supports adaptive and resilient interactions among cells, facilitating seamless integration and cooperation, including easy discovery and communication with new cells. It enables a GenAI-native system to dynamically respond to changing requirements and conditions, promoting scalability, resilience, and agility. 

\subsubsection*{Reproducible Infrastructure}

Immutable infrastructure involves deploying microservices as immutable entities to facilitate reproducibility. In case of GenAI-native cells, this principle needs to be relaxed to allow application logic and state to evolve at runtime, while retaining easy reproducibility.
To achieve this, the organic and self-evolving nature of GenAI-native cells must be captured and stored in a manner that allows for easy reconstruction, ensuring that the cloned cell retains the evolved capabilities of the original cell. 

It will be essential to minimize the degree of evolvability of such cells relative to their original base 'stem' cell, to facilitate easy reproducibility and restricting excessive variations, thereby preventing cells from becoming dysfunctional or "cancerous". Long-term evolution of cells to incorporate fundamentally new capabilities should still be handled in a coordinated manner, outside a live GenAI-native system.

\subsection{Programmability and Software Design Patterns}

Efficiently programming, configuring, and controlling GenAI-native systems requires new programming paradigms. New patterns and principles need to be developed and validated to address challenges such as unreliable outputs (i.e., non-deterministic, lower quality), the evolvability of assets and data structures and future human-agent codesign. 

Principles from existing specialized domains may be adopted, including modular code design, or probabilistic and live programming. However, a more radical approach, such as the creation of new programming languages or paradigms, may be required. A detailed analysis of software and programmability patterns in the context of designing and managing GenAI-native systems is beyond the scope of this paper.

\section{The GenAI-Native Software Stack}
\label{sec:sw_stack}

In this section we propose possible extensions to the traditional cloud-native software stack~\cite{aws_cloud_native, netesanyi2024cloudnative} and current agentic cloud platforms~\cite{aws-bedrock-agents-2025, azure-ai-foundry-agent-2025, gcp-vertex-ai-agent-2025} to build, manage, and run future GenAI-native applications and systems.

\subsubsection*{Infrastructure Layer} 
Minimal infrastructure requirements to efficiently support GenAI-native systems include the availability of specialized processing, networking and storage hardware for training, fine-tuning, or running (batch) inferencing jobs for small and large GenAI models. In particular, energy and runtime efficiency will be crucial, as future GenAI-native systems will include assets that constantly remain active, requiring sustainable and cost-effective operation. For securely deploying GenAI assets, support for confidential runtime environments may be required.

\subsubsection*{Provisioning Layer}
This layer facilitates the automated allocation and configuration of a low-level runtime environment for deploying applications and services. The self-reliant, evolvable aspects of GenAI-native applications may require revisiting existing frameworks and tools to better support their adaptive and organic nature, particularly when deployed in heterogeneous and distributed environments.

\subsubsection*{Runtime Layer}
This layer manages low-level virtual block and object storage capabilities, container runtime management, and virtual networking functionalities. The evolvability of GenAI-native systems requires highly efficient checkpointing capabilities to ensure seamless and rapid reproducibility when evolved GenAI-native cells need to be recreated elsewhere. Additionally, flexibly programmable organic virtual networks will be required to securely and dynamically intertwine multiple GenAI-native systems. Features such as cognitive firewalling, sandboxing, and cybersecurity may already be partially provided within this layer. 

\subsubsection*{Orchestration and Management Layer}

This layer is responsible for the low-level scheduling and orchestration of microservices, including service discovery, coordination, and management. In GenAI-native systems, we anticipate an increased need for handling more dynamic workloads, such as asynchronous worker agents proactively scheduling and executing new tasks. Service discovery and coordination frameworks should natively support organic and potentially federated service communication patterns. 

Furthermore, we expect a significant increase in additional service management frameworks to better accommodate the requirements of GenAI-native applications. Examples include service-level cognitive firewalling, sandboxing, and coordination across GenAI-native cells or subsystems. Additionally, managed services to facilitate systematic checkpointing of evolved logic and state, as well as native support for flexible, self-reliant agencies, will be essential.

\subsubsection*{Application Definition and Development Layer}

We envision a new hybrid programming model that seamlessly blends passive core logic with active GenAI-native assets. This model, possibly partly inspired by live programming paradigms, should prioritize the easy evolvability of code and data structures as first-class citizens. Additionally, such a hybrid, evolvable programming model should inherently support human-agent codesign, enabling humans and agents to collaborate on common artifacts and services.

Databases and storage services should facilitate versioned checkpointing of code and data to support evolvability, rollback, and reproducibility. Communication and streaming frameworks should enhance support for asynchronous conversational communication, including UCI and metadata.  

Source code management frameworks, along with CI/CD frameworks, should allow running GenAI-native systems to contribute changes in an easy and safe manner. Application and service definition frameworks will become increasingly important to easily specify, manage and control the intent of all GenAI-native assets and subsystems. This will allow other components to verify, constrain, and penalize the behavior of self-reliant and evolvable GenAI assets. Existing artifact registries and hubs should support more organic artifacts. 
Finally, third-party applications should become GenAI-native.

\subsubsection*{Observability and Analysis Tools}
Due to the organic and unpredictable nature of GenAI-native systems, they must be designed with robust observability, auditing, and analysis frameworks from the ground up. These frameworks are essential for continuously monitoring, evaluating, and improving the efficiency and effectiveness of all components and interactions across all layers, akin to human organizational paradigms. Existing GenAI logging and observability tools must be enhanced to include more advanced cognitive capabilities and implement effective, comprehensive, and actionable mechanisms to measure and address utility-based sufficiency criteria.

\section{Impact}
\label{sec:impact}

In this section, we provide a brief introduction to the high-level implications and impact of implementing the proposed guidelines and patterns for developing reliable and evolvable GenAI-native software systems. This paper offers only a high-level overview; more comprehensive analyses will be required for each of the aspects discussed.

\subsection{Technical Aspects}

\subsubsection*{Computational Overhead}
Despite some notable exceptions \cite{wang2024leveraging, xu2024llm}, the integration of GenAI into software systems can lead to substantial runtime and performance overhead compared to traditional methods. Moreover, GenAI currently demands a considerable resource footprint, posing significant economic and environmental challenges. When future software systems will evolve to become \textit{GenAI-native} and become increasingly self-reliant, these challenges are likely to intensify.

In this paper, we proposed several guidelines and patterns to minimize unnecessary reliance on GenAI processing. Additionally, it will be essential to identify the optimal balance for deploying autonomous, self-reliant agents, incorporating cost-benefit analysis and return-on-investment criteria.

\subsubsection*{Interoperability and Scalability}

The transition to GenAI-native systems is poised to introduce significant operational challenges, particularly in ensuring seamless interoperability. These systems must effectively communicate with both traditional legacy systems and other emerging GenAI-native systems. Furthermore, scalability will remain a critical concern, as initial GenAI-native implementations will encounter limitations when handling increased workloads and expanding functionalities. 

Consequently, the adoption of GenAI-native systems is expected to be incremental, progressively integrating advanced cognitive and autonomous capabilities to enhance system capabilities. Establishing industry standards for GenAI-native systems and communication protocols can facilitate smoother integration and interoperability. In addition, designing systems with modular architecture can help in scaling and integrating new GenAI capabilities without disrupting existing functionalities.

\subsubsection*{Privacy and Security}

Future GenAI-native systems must technically address critical privacy and security concerns, such as ensuring data protection, enforcing secure access and authorization mechanisms, and maintaining compliance with regulations. Although we have already outlined several high-level guidelines and patterns, these considerations may inevitably impact the adoption rate of advanced GenAI-native capabilities within existing application architectures.


\subsection{User Experience and Adoption}

\subsubsection*{Learning Curve}
The transition to developing and managing GenAI-empowered software assets presents adoption challenges for software engineers throughout the entire lifecycle, including roles such as architects, designers, developers, testers, and maintenance personnel. Although the design principles and patterns proposed in this paper aim to facilitate this transition, they themselves introduce an learning curve that will require training and experience.

To mitigate these challenges and ease the transition, it is essential to develop user-friendly tools and frameworks. Examples include tools for configuring and managing programmable routers, implementing reflective communication and processing mechanisms, and creating resilience mechanisms tailored to specific use cases or domains. 

Additionally, frameworks should be developed to support the easy development and management of, for example, fully functional GenAI-native service cells, the creation of unified conversational interfaces (e.g., the Agora protocol~\cite{marro2024scalablecommunicationprotocolnetworks}), and cognitive workflow optimizers.
Finally, existing or novel programming paradigms should be integrated and developed to facilitate the programming of assets with predictability and reliability issues, as well as easy programming with malleable, evolving data.

\subsubsection*{User Experience}
The adoption of GenAI and agentic services is poised to significantly impact user experience and interaction. Currently, chat-based interfaces are already partly replacing traditional information browsing and retrieval paradigms. Numerous new AI browsers are entering the market, aiming to redefine existing browsing experiences and user interactions~\cite{newton2025aibrowserwars}. In addition, GenAI-based services promise enhanced personalization and facilitate end-user programming paradigms~\cite{litt2023malleable}. Moreover, the emergence of new devices and gadgets is targeted at providing more intuitive interactions with these services. 

These innovations will also require a shift in user mentality, yet GenAI's inherent capabilities can hopefully help facilitate this transition. However, ensuring dependability and building trust in these new interfaces and services remain paramount to their successful adoption.

\subsubsection*{Societal and Ethical Considerations}
GenAI-empowered applications and tools are already significantly influencing people's work dynamics, roles, and daily tasks. Given this societal impact, it is crucial to ensure the responsible use of GenAI in future software systems, including implementing bias mitigation strategies. This paper emphasizes the importance of balancing GenAI-based processing with traditional techniques, with many proposed guidelines and patterns aimed at achieving this balance. 

Furthermore, the primary objective of self-reliant GenAI agents should be to enhance cognitive workflows by reducing their involvement in critical paths. This involves thoroughly examining all generated automation and optimization solutions, potentially with human oversight, to ensure reliability and effectiveness.

\subsection{Economic and Business Implications}

\subsubsection*{Economic Implications}

Developing and managing GenAI-native software systems involves several critical economic considerations. One significant challenge is the potential runtime overhead and increased computational footprint introduced by GenAI technologies. Additionally, the development and maintenance costs for GenAI-native systems remain uncertain, particularly in the early stages when suitable tools, frameworks, and best practices will still be under development and refinement.

As discussed, GenAI assets may exhibit variations in behavior when the underlying LLM or GenAI technology is replaced, leading to additional overhead and potential convergence issues that must be addressed. Furthermore, the return on investment (ROI) for deploying self-reliant autonomous agents and self-evolving or personalized software systems requires careful monitoring and control to ensure economic viability.

\subsubsection*{Operational Efficiency}
One of the primary focus areas of this paper is to systematically enhance operational efficiency within GenAI-native systems while simultaneously improving overall reliability. 
To achieve profitability, continuous productivity enhancements and process optimizations are vital. Consequently, the guidelines and patterns discussed are specifically designed to support these objectives, ensuring that GenAI-native systems operate efficiently, alongside with improving overall reliability. 

\subsubsection*{Market Impact}
The transition to GenAI-native systems may significantly influence market dynamics. In the short term, early adopters may gain a competitive edge over traditional, more rigid solutions. In the long term, the nature and dynamics of services are expected to undergo radical transformations. For instance, existing business-to-business (B2B) and business-to-consumer (B2C) models may evolve into business-to-agent (B2A) and agent-to-agent models~\cite{ladsariya2025agenttoagent}, where autonomous agents and services negotiate and interact on behalf of humans or organizations, leading to potentially volatile market dynamics.

GenAI-native web services are also likely to engage with each other more proactively and dynamically compared to their traditional, passive counterparts. As a result, GenAI-native solutions will continuously need to demonstrate and maintain their value in the face of constantly evolving competing alternatives, potentially causing major ripples into the existing market.

\subsection{Legal and Regulatory Aspects}

\subsubsection*{Compliance}

Future GenAI-native systems must comply with evolving laws and regulations, including intellectual property rights. This paper briefly discussed the implementation of a robust policy language and framework to meet these requirements and facilitate regular audits by external parties. Particularly for self-evolving systems, ensuring continuous compliance is crucial yet challenging. It is imperative to incorporate strong fail-safe mechanisms to address potential failures and maintain adherence to regulatory standards.

\subsubsection*{Liability}
As GenAI-native systems become increasingly autonomous, it is crucial to establish accountability for AI decisions, generated responses, and actions. Preventing these systems from triggering harmful actions or leading humans to make incorrect decisions due to erroneous outputs is essential. Furthermore, the liability of human organizations and enterprises deploying and managing such systems must be clearly defined.

The EU AI Act~\cite{EU_AI_Act_2024} already sets forth regulations concerning AI and autonomous systems. However, the classification of agentic AI systems as high-risk, requiring full compliance, remains ambiguous~\cite{Lexology_Agentic_AI_2025}. For future GenAI-native systems, which will integrate both traditional and agentic subsystems, it is vital to delineate the boundaries and interactions between these processing types. Although this might slow the adoption of advanced self-reliant GenAI-native systems, ensuring user safety remains key.

\subsubsection*{Data Governance}

Finally, GenAI-native systems will also need to establish and enforce robust data governance protocols, governing data retention, usage, and sharing, both internally and across multiple GenAI-native systems. Beyond traditional policies, it is imperative to define explicit guidelines regarding the utilization of user-specific data and interactions, particularly when enhancing, customizing, or evolving the functionality and operational excellence of such systems and assets. This paper outlines several privacy and security guidelines and patterns. Heightened attention will be required in scenarios involving highly cognitive self-reliant and self-evolving operations.

\vspace{3mm}
\section {Conclusions and Future Work}

In this paper, we advocated for novel design principles and paradigms for building robust and adaptive GenAI-native systems, emphasizing the need for such approach. We discussed core design principles built around five key pillars: reliability, excellence, evolvability, self-reliance, and assurance. Drawing from historical and human perspectives, and centered around several example use cases, we proposed new guidelines, which were crystallized into foundational patterns such as the GenAI-native cell, programmable router, and organic substrate. These aim to reduce the inherent complexity of GenAI-native systems while retaining its potential. We also briefly discussed the GenAI-native software stack, and touched upon the possible impact and implications of future GenAI-native systems from a technical, user adoption, economical, and legal perspective. 

Correctly implemented GenAI-native systems should be robust and flexible against both expected and unexpected issues, even beyond those caused by GenAI, with built-in mechanisms for self-reliant evolution. However, poor decision-making can lead to system failure, necessitating shutdown or overhaul. 
Note that we did not cover all aspects, such as the interplay with robotics and cyber-physical systems. Additionally, most ideas require further validation through experimentation and real-life application, to better understand the impact and benefits of the proposed patterns. 

Finally, while current GenAI technologies may not yet fully support the implementation of many ideas presented in this paper, we expect significant advancements in the coming years. We hope this paper will inspire various communities and offer a theoretical framework for developing robust and adaptive future GenAI-native systems.

\twocolumn
\bibliographystyle{ACM-Reference-Format}
\balance
\bibliography{references}

\end{document}